\documentstyle[epsf]{mn}  
\begin{document}   
\title[The Effects of Spatial Correlations on Merger Trees]{The
Effects of Spatial Correlations on Merger Trees of Dark   
Matter Haloes}    
\author[M. Nagashima \& N. Gouda]{Masahiro   
Nagashima\thanks{Research Fellow of the Japan Society for the   
Promotion of Science} and Naoteru Gouda\\    
Department of Earth and Space Science, Graduate School of Science,\\   
Osaka University, Toyonaka, Osaka 560, Japan;\\    
Email: masa, gouda@vega.ess.sci.osaka-u.ac.jp}   
\maketitle   
   
\begin{abstract}   
The effects of spatial correlations of density fluctuations on merger
histories of dark matter haloes (so-called `{\it merger trees}') are
analysed.  We compare the mass functions of dark haloes derived by a
new method for calculating merger trees, that proposed by Rodrigues \&
Thomas (RT), with those given by other methods such as the Block
model, the Press-Schechter formula and our own formula in which the
mass functions are analytically expressed in a way that takes into
consideration the spatial correlations.  It is found that the mass
functions given by the new method are well fit by those given by our
formula.  We believe that new method (RT) {\it naturally} and
correctly takes into account the spatial correlations of the density
fluctuations due to a calculated, grid-based realisation of the
density fluctuations and so is very useful for estimating the merger
tree accurately in a way that takes into consideration spatial
correlations.
   
Moreover, by applying our formula, we present an analytic expression
which reproduces the mass function derived by the Block model.  We
therefore show clearly why and how the mass functions given by the new
method and the Block model are different from each other.
Furthermore, we note that the construction of merger trees is
sensitive to the criterion of collapse and merging of overlapped
haloes in cases in which two or more haloes happen to overlap.  In
fact, it is shown that the mass function is very much affected when
the criterion of overlapping is changed.
\end{abstract}   
   

\section{INTRODUCTION}   

Recent observations by the HST and the Keck telescope have rapidly
increased the available numbers of observations of faint, high
redshift galaxies, providing us with significant new information about
the birth and evolution of galaxies.  In order to understand the
significance of these observations, it is very important to
theoretically understand the physical processes underlying galaxy
formation.
   
In our standard understanding, it is considered that dark matter
dominates in our Universe and that galaxies and clusters of galaxies
have formed by the gravitational growth of initial small density
fluctuations. The fluctuations of dark matter have collapsed and
virialised by self-gravitational instability into objects which are
called `dark matter haloes' or `dark haloes'.  The larger haloes are
generally considered to have formed hierarchically by clustering of
smaller haloes (it is so called `hierarchical clustering').
   
The baryonic gas also has collapsed and virialised following the
collapse of the dark matter.  Furthermore, in the process of galaxy
formation, baryonic gas must dissipate the internal energy by
radiative cooling and shrink because galaxies are much denser systems
than the virialised dark haloes and stars must be formed inside the
cold and dense gaseous systems.  In analysing the formation and
evolution of the galaxies, we must treat various physical processes
over a large dynamic range from $\sim 1-10^{2}M_{\sun}$ (star
formation, heating processes of gases from supernovae, dynamical and
chemical evolution of gases, etc.) through $\sim 10^{6-12}M_{\sun}$
(mergers of galaxies, tidal interactions, etc.)  to $\sim
10^{13-16}M_{\sun}$ (clusters of galaxies, large scale structure of
the Universe, etc.).  So it is difficult to attack the problem of
formation and evolution of galaxies in a way that connects all of the
above complicated physical processes.
   
One method of analysing galaxy formation is numerical simulations
(e.g., Navarro \& White 1993, Katz \& Gunn 1991, Katz 1992, Steinmetz
\& Muller 1994, Evrard, Summers \& Davis 1994) which directly pursue
the gravitational growth of dark matter and thermal processes of
gaseous systems.  The advantage of simulation is that we can trace the
complicated processes of the systems quantitatively. However, it is
impossible to deal with the very wide range of the mass scales
simultaneously in the current limited ability of computers.
Furthermore, since the CPU time is heavily consumed in the
simulations, it is difficult to analyse galaxy formation statistically
by pursuing many samples of the system with different initial
conditions and parameters in the physical processes. So we can say
that the problem of galaxy formation has never been resolved
completely by numerical simulations without any uncertainty.
   
On the other hand, there is another approach to solving galaxy
formation, that is, the semi-analytical approach, which has been
developing.  Some works in this approach are briefly reviewed below.
The pioneer works in this approach are, for example, Rees \& Ostriker
(1977) and White \& Rees (1978).  Rees \& Ostriker remarked upon the
importance of the dissipational process of baryonic gases through
radiative cooling.  They asserted that a dense object like a galaxy
must have a cooling time scale $\tau_{cool}$ shorter than the
dynamical time scale $\tau_{dyn}$ of the system at the virialised
stage because the object must dissipate the energy effectively in
order to become a more dense system and make stars in the cold and
dense gases.  Therefore, one of the criteria for a virialised object
to become a galaxy in the future is $\tau_{cool} < \tau_{dyn}$ at the
virialised stage.  Then Rees \& Ostriker showed in the cooling diagram
that the range $\tau_{cool}<\tau_{dyn}$ corresponds to the
characteristic mass range of galaxies, {\it i.e.},
$10^{8}M_{\sun}\sim 10^{12}M_{\sun}$.
   
White \& Rees (1978) showed the luminosity function of galaxies by
using the Press-Schechter formula (Press \& Schechter 1974;hereafter
PS, {\it see} below) in which the mass function of dark haloes is
analytically estimated.  Furthermore, White \& Frenk (1991) extended
this approach to consider the {\it merging process} of dark haloes
approximately by using an extended PS formula (Bower 1991). In this
formula they take into account the conditional probability of finding
a region with mass $M_{1}$ at redshift $z_{1}$ which is also included
in a region with mass $M_{2}$ at $z_{2}$.  They calculate the present
stellar and gas abundance by tracing the merger process of the dark
haloes from an initial time till the present time in each mass scale
of the halo on the assumption that there is a relation between the
halo mass, the gas cooling rate, the accretion rate of gas onto the
halo and the star formation rate.  The extension of the PS formula is
also discussed by Lacey \& Cole (1993).  By using the peak formula
(Peacock \& Heavens 1985, Bardeen et al. 1986; hereafter BBKS), Lacey
\& Silk (1991, 1993) also investigate the time scale of the collapse
of galaxy groups after the galaxies have collapsed on the assumption
that the mass ratio of groups to galaxies is constant.
   
However, the status of the gaseous systems and formations of stars in
each dark halo at an epoch depend on the thermal history of the
gaseous systems, merging of the galaxies and the merging history of
the halo.  So it is very important to know the merging history of each
dark halo at an epoch in order to evaluate the status of the gaseous
systems and the formation rate of stars in the halo at that epoch.
Kauffman \& White (1993; hereafter KW) constructed a {\it merger tree}
that expresses the merging history of the dark haloes by considering
progenitors of each halo at every time by using a Monte Carlo method
of the extended PS formula.  Kauffman, White \& Guiderdoni (1993) then
calculated the statistical properties of galaxies in their model.
Another approach to construction of a merger tree is the Block model,
which was developed by Cole \& Kaiser (1988).  The Block model takes
the Monte Carlo procedure described as follows: First of all, the
density contrast is assigned to the largest block with mass
$M_{0}(\sim 10^{16}M_{\sun})$ which has the variance of the density
contrast, $\sigma^{2}(M_{0})$.  Next, the largest block is divided
into two blocks with the same mass $M_1(=M_0/2)$. The additional
positive density fluctuation generated by a random Gaussian
distribution with variance $\sigma^{2}(M_{1})-\sigma^{2}(M_{0})$ is
assigned to one of the two divided blocks with mass $M_{1}(=M_{0}/2)$
and a negative fluctuation with the same absolute value of the
amplitude as the positive one is assigned to the other divided block.
The same procedure is repeated for $M_1, M_2, \cdots$ down to the
smallest mass scale under consideration, thereby constructing the
merger tree.  Cole et al. (1994) also calculated the statistical
properties of formation and evolution of galaxies in their Block
model.
   
Incidentally, it has been pointed out that the PS formula has the
following crucial weak points.  The PS formula is derived as follows:
In an Einstein-de Sitter universe, spherical overdense regions
collapse and virialise when their linear density contrast reaches
$\delta_{c}=1.69$ (see, e.g., Peebles 1993).  Then, assuming that the
density fluctuations obey a random Gaussian distribution and that the
collapse of the haloes is spherically symmetric, we can get the volume
fraction of the regions of collapsed objects whose masses are larger
than the mass $M$, $f(\geq\delta_{c},M)$.  So, the region of the dark
haloes with mass scale $M$ is equal to $\partial f/\partial M$.  Thus
they proposed the formula for counting the number density of objects
of mass scale $M$.  However, only overdense regions were considered in
the analysis.  Even if the density contrast smoothed on the mass scale
$M$ is less than $\delta_c$, there is the case such that the density
contrast smoothed on the larger mass scale than $M$ is greater than
$\delta_c$.  We must consider this case to count exactly the number
density of the dark haloes.  This problem is called the
`cloud-in-cloud' problem.  Press \& Schechter (1974) {\it simply}
multiplied the number density by a `fudge factor' of 2 without a good
reason.
   
Peacock \& Heavens (1990) and Bond et al. (1991) attempted to solve
the cloud-in-cloud problem by using a peak formula.  They considered
the {\it upcrossing} probability that a density contrast which is
lower than $\delta_{c}$ with a smoothing scale $M$ exceeds
$\delta_{c}$ for the first time when increasing the smoothing scale.
In the case that the density field is smoothed with the sharp {\it
k}-space filter, they found that the factor of 2 introduced by PS is
correct.  The cloud-in-cloud problem is also solved for Poisson
fluctuations by Epstein (1983).
   
In the cloud-in-cloud problem, however, we must consider the {\it
spatial correlations of the density fluctuations} since objects have
non-zero size in reality.  Yano, Nagashima \& Gouda (1996; hereafter
YNG) analysed the cloud-in-cloud problem taking explicitly into
account the spatial correlation of the density fields.  They
explicitly introduced the two-point correlation function in the mass
function by using Jedamzik's formula (Jedamzik 1995) in which the
number density of the collapsed objects is given in the form of the
integral equation which is different from the PS formula.  It is also
proved by YNG that the results derived from the Jedamzik formula are
consistent with those derived from the PS formula on the assumption
that the collapse is spherically symmetric when the spatial
correlations are not taken into consideration.  However, YNG showed
that the spatial correlations greatly affect the mass function.
Therefore, we believe that the merger tree is also affected by the
spatial correlations whose effects have never explicitly been taken
into account in the KW method and the Block model.  So we believe it
is very important to analyse the effects of spatial correlations of
the density fields on the merger trees of dark haloes.
   
Recently, a new approach to construction of merger trees has been
proposed by Rodrigues \& Thomas (1995) (we call this the {\it Merging
Cell model} for convenience throughout this paper).  In their model,
the random Gaussian density fluctuation field is realised on spatial
grids as in constructing initial conditions of N-body
simulations, so it is expected that this model naturally includes
information about the spatial correlation.  Then, by finding the
region of the collapsed cells or blocks with each mass scale whose
density contrast $\delta$ is $\delta_{c}$, we can construct the merger
tree (see $\S 2.1$).  We expect that this model is more realistic and
useful for galaxy formation although spherical collapse and random
Gaussian density fluctuations are assumed.
   
In this paper, we show the mass functions of dark haloes by
calculating the Merging Cell model.  They are compared with the mass
functions given by the PS formula and the Block model.  We then
explicitly show that the main origin of differences between the mass
functions given by the different methods result from the effect of
spatial correlations by comparing with the mass functions derived by
YNG's formula (hereafter, the YNG formula).  We find that the mass
functions given by the YNG formula in taking explicitly into account
the spatial correlations are consistent with those given by the
Merging Cell model and so the Merging Cell model correctly includes
the effects of the spatial correlations.  We believe that this effect
is very important for calculating the merger tree of the dark haloes.
Furthermore, by applying the Jedamzik formula, we present an
analytical expression of the mass function derived from the Block
model and show quantitatively why and how the mass function given by
the Block model is different from those derived from the PS formula
and the Merging Cell model.
   
In the Merging Cell model, some haloes happen to overlap since the
non-zero size of the haloes is considered explicitly. So, we must
consider the criterion of collapse and merging for the overlapped
haloes.  We also show how the criterion affects the mass function in
the Merging Cell model.

In \S 2, the Merging Cell model and the Block model are reviewed
briefly.  In \S 3, we give the analytical formulae for estimating the
mass functions by using the Press-Schechter formula, the Jedamzik
formula and the YNG formula.  In \S 4, it is shown that the Merging
Cell model is consistent with the mass function which is derived by
the analytical formula in which the spatial correlations are taken
into account.  The mass function given by using the Jedamzik formalism
which reproduces those given by the Block model is also presented.
Furthermore, we also show how the mass function is changed if the
overlapping criterion is changed.  We devote \S 5 to conclusions and
discussions.
   
\section{MODELS OF MERGER TREES}   
   
In this section, we briefly review the Merging Cell model (hereafter
MCM) and the Block model.
   
\subsection{Merging Cell model}   
   
We briefly review the MCM according to the procedure and the notations 
shown in Rodrigues \& Thomas (1995). 
   
First, the random Gaussian density field is realised in a periodic
cubical box of side $L$.  In the random Gaussian distribution, Fourier
mode of density contrast $\delta(=(\rho-\bar\rho)/\bar\rho, \rho$ is
density and $\bar\rho$ is the mean density of the universe) obeys the
following probability for its amplitude and phase(BBKS),
\begin{equation}   
P(\vert\delta_{\bf k}\vert,\phi_{\bf k})d\vert\delta_{\bf k}\vert   
d\phi_{\bf k}=\frac{2\vert\delta_{\bf k}\vert}{P(k)}\exp\left\{-\frac{   
\vert\delta_{\bf k}\vert^{2}}{P(k)}\right\}d\vert\delta_{\bf   
k}\vert\frac{d\phi_{\bf k}}{2\pi},   
\end{equation}   
where $\phi_{\bf k}$ is the random phase of $\delta_{\bf k}$, 
$\delta_{\bf k}=|\delta_{\bf k}|\exp{(i\phi_{\bf k})}$ and $P(k)$ is 
the power spectrum $\langle|\delta_{\bf k}|^{2}\rangle$, where the 
angle brackets mean the ensemble average of the universe.  Then, the 
density contrast at each grid (`cell') is given by Fourier 
transform, 
\begin{equation}   
\delta({\bf x})=\frac{V}{(2\pi)^{3}}\int_{0}^{k_{c}}\delta_{\bf   
k}e^{i\bf k\cdot x}d^{3}k,   
\end{equation}   
where $k_{c}$ is the cut-off wavenumber.     
   
Next, averaging the density fluctuations within cubical {\it blocks}
of side 2, 4, \ldots, L, the fluctuations of the various smoothing
levels are constructed.  At each smoothing level, displacing the
smoothing grids by half a blocklength in each direction of each axis,
eight sets of {\it overlapping grids} are constructed in order to
reflect the position of density peaks approximately (see Fig.1).
   
Then, the density fluctuations within blocks and cells are combined
into a single list and ordered in decreasing density.  The
fluctuations are investigated from the top of the list.  It is decided
by the following rules whether each block or cell can collapse.  Note
terminology that {\it halo} is a block or cell which has already
collapsed, and an {\it investigating} region is a block or cell whose
linear density contrast is just equal to $\delta_c$ at the reference
time.  We investigate whether or not an investigating region can
collapse at that time according to the following rules.
\begin{enumerate}   
\renewcommand{\labelenumi}{(\alph{enumi})}
\item If an investigating region includes no haloes, the investigating 
region({\it block} or {\it cell}) can collapse and can be identified 
as a new halo. 
\item When an investigating region includes a part of a halo, if the
{\it overlapping region} has a larger mass than the minimum of the
masses of the halo and the investigating region, then the
investigating region can collapse (see Fig.2).  This is the criterion
of collapse of the investigating region and merging for the overlapped
haloes. We call this criterion {\it the overlapping criterion} in this
paper.
\item In the case of (b), if the investigating region has two haloes
whose overlapped region within the investigating region is more than a
half of the region in each halo, the region cannot collapse.  We set
this condition in order to avoid long filamentary objects.  This is
the criterion for linking of haloes.
\end{enumerate}   
   
These criteria are those chosen by Rodrigues \& Thomas. We will change
the overlapping criterion (b) later and see how the mass function is
changed.  The condition (c), the linking condition, prevents the
growth of filamentary structures because galaxies that we observe are
not filamentary.  However, we should note that the filament structures
of dark matter certainly appear as shown in numerical
simulations. Then the criterion for linking is just an assumption.
   
It must be noted here that in the MCM, because overlapping grids are
used, the mass spectrum is close to continuous, rather than restricted
to powers of two as in the Block model.
 	   
\begin{figure}
\epsfxsize=8cm
\epsfbox{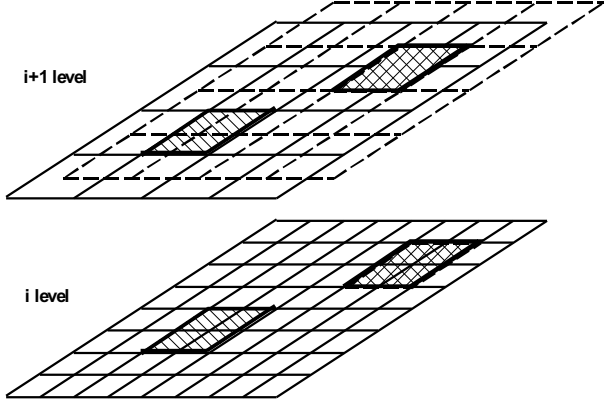}
\caption{Scheme of averaging density fluctuations demonstrated in 2
dimensions (after Rodrigues \& Thomas 1995).  Lower grid represents
$i$-th level.  Upper grids represent $i+1$-th level with one of the
overlapping grids displaced by half a blocklength.  The hatched region
in the lower grid is averaged to make a block at the $i+1$-th level
and indicated as the region marked as a thick square in the upper
grid.  Blocks in the offset grid are averaged from the lower level in
the same way.}
\end{figure}

\begin{figure}
\epsfxsize=8cm
\epsfbox{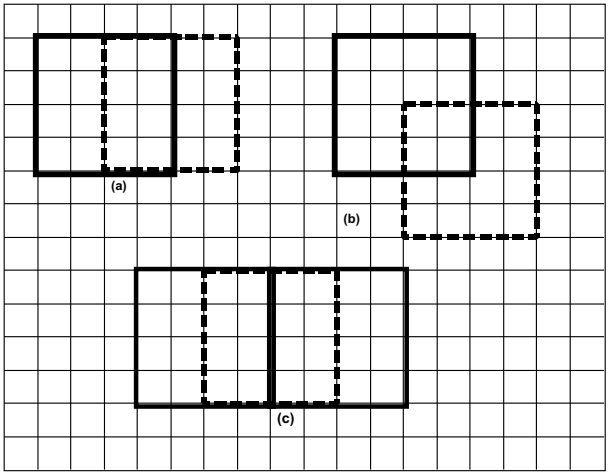}
\caption{The overlapping criterion.  The thick squares show a halo and
the dashed squares show an investigating region.  (a)The block can
collapse and is merged into the halo because the overlapping region is
larger than half of the lesser of the regions of the halo and the
investigating region.  (b)The block cannot collapse because the
overlapping region is not larger than half of the lesser of the
regions of the halo and the investigating region.  (c)The linking
condition.  In the case shown here, the block cannot collapse because
we would like to prevent the growth of the long filamentary structure
(see text).  Note that it is equivalent to consider either regions or
masses for the overlapping criterion.  }
\end{figure}

\subsection{Block model}   
   
In this subsection, the Block model is briefly reviewed according to
Cole \& Kaiser (1988).  Now, we consider a large block with mass
$M_{0}$($\sim 10^{16}M_{\sun}$), divide this into two, and realise
blocks with various mass scales by successively dividing smaller and
smaller blocks.  We obtain a set of blocks of discrete masses,
$M_{i}=M_{0}/2^{i}(i= 1, 2, \cdots, N)$ where $N$ is an integer and
$N$ is 27 in our calculation as shown later.
   
Here $\sigma(M_{i})=\sigma_{i}$ means the standard deviation of the
density fluctuations smoothed on a mass scale $M_{i}$.  When the power
spectrum is written in the scale-free form $P(k)\propto k^{n}$,
the standard deviation is given by
\begin{eqnarray}   
\sigma_{i}^{2}&=&\frac{4\pi
	V}{(2\pi)^3}\int_{0}^{\infty}W_{i}^{2}(k)P(k)k^{2}dk
	\nonumber\\
	&=&\left(\frac{M_{i}}{M_{*}}\right)^{-\frac{3+n}{3}}\label{eqn:vari},
\end{eqnarray}   
where $M_{*}$ is the mass on which the variance is unity, and
$W_{i}(k)$ is the window function.  We define the quantities
$\Sigma_{i}$ as follows:
\begin{eqnarray}   
\Sigma_{0}&=&\sigma_{0},\\
\Sigma_{i}^{2}&=&\sigma_{i}^{2}-\sigma_{i-1}^{2},~~~i\geq
1\label{eqn:sharp}.
\end{eqnarray}   
   
We generate the random density fluctuations on each block as follows:  
First of all, a density contrast generated by the random Gaussian  
distribution with the variance $\Sigma_{0}$ is assigned to the largest  
block.  This corresponds to the density contrast smoothed on the mass  
scale $M_{0}$.  Then the positive random variable generated by the  
random Gaussian distribution with the variance $\Sigma_{1}$ is added  
to one of the two divided blocks with the mass $M_{1}=M_0/2$ and a   
negative one with the same absolute value of the amplitude is added to  
the other divided block. Repeating this procedure from $i=0$ until  
$i=N$, we obtain a tree of density contrasts of the blocks with  
various mass scales.  The condition of collapse for a block at the  
redshift $z$ is that the density contrast of the block equals  
$\delta_{c}=1.69(1+z)$ ($z$; redshift). In this way, we can construct  
the merger tree of the dark haloes.  
    
\section{ANALYTIC APPROACH TO MASS FUNCTIONS}   
   
In this section, we present the PS formula, the Jedamzik formula and
also the mass functions derived by our formula (YNG formula) which
explicitly includes the effects of the spatial correlations.
   
\subsection{Press-Schechter formula}   
The probability of finding the region whose linear density contrast
smoothed on the mass scale $M$, $\delta_{M}$, is greater than or equal
to $\delta_{c}$ is assumed to be expressed by the random Gaussian
distribution given by
\begin{equation}   
	f(\geq\delta_{c},M)=\frac{1}{\sqrt{2\pi}\sigma(M)}   
	\int_{\delta_{c}}^{\infty}\exp\left(-\frac{\delta^{2}}{2\sigma^{2}   
	(M)}\right)d\delta.   
\end{equation}   
This probability corresponds to the ratio of the volume of the region
above $\delta_{c}$ in the density contrast smoothed on the mass scale
$M$ to the total volume (in a fair sample of the Universe).  Therefore,
the difference between $f(\geq\delta_{c},M)$ and
$f(\geq\delta_{c},M+dM)$ represents the volume of the region for which
$\delta_{M}=\delta_{c}$ precisely.  The density contrast of an {\it
isolated} collapsed object is precisely equal to $\delta_c$ because an
object with $\delta>\delta_c$ would be eventually counted as an
object of larger mass scale.  The volume of each object with mass
scale $M$ is $M/{\bar\rho}$. Then we obtain the following relation,
\begin{equation}   
  \frac{Mn(M)}{\bar\rho}dM=-\frac{\partial   
f(\geq\delta_{c},M)}{\partial M}dM,   
\end{equation}      
where $n(M)$ means the number density of objects with mass $M$,   
that is, the mass function.   
However, the underdense regions are not considered in the above equation.   
Hence, Press and Schechter {\it simply} multiply the number density by a   
factor of 2,   
\begin{equation}   
  \frac{Mn(M)}{\bar\rho}dM=-{\it 2}\frac{\partial f(>\delta_{c},M)}   
  {\partial M}dM.   
\end{equation}      
This factor of 2 has long been noted as a weak point in the PS
formula (the so-called `cloud-in-cloud' problem).  Peacock \&
Heavens (1990) and Bond et al. (1991) proposed a solution to this
problem by taking account of the probability $P_{up}$ that subsequent
filtering an larger scales might result in having $\delta >
\delta_{c}$ at some point, even when at smaller filters, $\delta <
\delta_{c}$ at the same point.  They found that the factor of 2 in the
PS formula could be correct only by using the sharp {\it k}-space
filter given by
\begin{equation}   
  W_{i}(k)=\left\{   
  \begin{array}{ll}   
    1,&\quad k\leq k_{c}(M_{i})\\   
    0,&\quad k>k_{c}(M_{i}).   
  \end{array}\right.   
\end{equation}

\subsection{Jedamzik formula}   
Jedamzik (1995) proposed another approach to the cloud-in-cloud problem.   
   
Now, we consider the regions whose smoothed linear density contrasts
on the mass scale $M_1$ are above $\delta_{c}$.  Each region must be
included in an isolated collapsed object with mass $M_{2}\geq M_{1}$.
Therefore, we obtain the following equation,
\begin{equation}   
  f(\geq\delta_{c},M_{1})=\int_{M_{1}}^{\infty}P(M_{1},M_{2})\frac{M_{2}}   
  {\bar\rho}n(M_{2})dM_{2},   
\end{equation}   
where $P(M_{1},M_{2})$ means the conditional probability of finding a
region of mass scale $M_1$ in which $\delta_{M1}$ is greater than or
equal to $\delta_c$, provided it is included in an isolated overdense
region of mass scale $M_2$.  By `the Jedamzik formula' we mean the
procedure in which the mass functions are estimated by solving
eq.(10).  If $P(M_{1},M_{2})$ is given by the conditional probability
$p(\delta_{M_{1}}\geq\delta_{c}|\delta_{M_{2}}=\delta_{c})$, $P(M_1,
M_2)$ is written as follows by using Bayes' Theorem,
\begin{eqnarray}   
P(M_{1},M_{2})&=&p(\delta_{M_{1}}\geq\delta_{c}|\delta_{M_{2}}=\delta_{c})
	\nonumber\\
	&=&\frac{p(\delta_{M_{1}}\geq\delta_{c},\delta_{M_{2}}=\delta_{c})}
	{p(\delta_{M_{2}}=\delta_{c})}\nonumber\\
	&=&\frac{1}{\sqrt{2\pi}\sigma_{\mbox{\scriptsize sub}}}\int
	_{\delta_{c}}^{\infty}\exp\left\{\frac{1}{2}\frac{(\delta_{M_{1}}-
	\delta_{c})^{2}}{\sigma^{2}_{\mbox{{\scriptsize
	sub}}}}\right\}d\delta_{M_{1}} \nonumber\\
	&=&\frac{1}{2}\label{eqn:jed},\\ \sigma^{2}_{\mbox{\scriptsize
	sub}}&=&\sigma^{2}(M_{1})-\sigma^{2}(M_{2}) \label{eqn:sub}
\end{eqnarray}   
where we use the sharp {\it k}-space filter (see YNG).  Thus we can
obtain the PS formula, naturally including the factor of 2 as can be
seen from eqs.(10) and (\ref{eqn:jed}).
   
However, it is insufficient for more realistic estimation of the mass  
function to use eq.(\ref{eqn:jed}) because it is necessary to consider  
the spatial correlation of the density fluctuations due to the finite  
size of the objects.  Therefore, we must consider the probability  
$P(r,M_{1},M_{2})$ of finding $\delta_{M_{1}}\geq\delta_{c}$ at a  
distance $r$ from the centre of an isolated object of mass scale  
$M_{2}$.  Then we can get the probability $P(M_1, M_2)$ by spatially  
averaging $P(r,M_{1},M_{2})$.  
   
Because we believe that the isolated collapsed objects are formed around  
density peaks, the constraints to obtain the above probability  
$P(r,M_{1},M_{2})$ are given as follows:  
\begin{enumerate}   
\item The linear density contrast, $\delta_{M_{2}}$, of the larger
mass scale $M_{2}$, should be equal to $\delta_{c}=1.69$ at the centre
of the object.
\item Objects of the mass scale $M_{2}$ must contain a maximum peak of   
the density field, {\it i.e.}, the first derivative of the density   
contrast $\nabla\delta_{M_{2}}$ must be equal to $0$ and each diagonal   
component of the diagonalized Hessian matrix $\zeta$ of the second   
derivatives must be less than $0$ at the centre of the object.   
\item The density contrast of the smaller mass scale $M_{1}(\leq  
M_{2})$ which collapsed and is included in an object of mass scale  
$M_{2}$ must satisfy the condition $\delta_{M_{1}}\geq\delta_{c}$ at  
distance $r$ from the centre of the larger object.  
\end{enumerate}   
The probability which we get from the above conditions is:   
\begin{eqnarray}   
\lefteqn{P(r,M_{1},M_{2}|\mbox{peak})=P(\delta_{M_{1}}\geq\delta_{c}|
	\delta_{M_{2}}, \mbox{peak})}\nonumber\\
	&&=\sqrt{\frac{1-\gamma^{2}}{2\pi(1-\epsilon^2-\mu^{2}-\gamma^{2}+
	2\epsilon\mu\gamma)}}\nonumber\\
	&&\qquad\qquad\times\frac{\int_{0}^{\infty}dx~f(x)\int_{\nu_{1c}}
	^{\infty}d\nu_{1}\exp(-\frac{Q_{a}+Q_{b}}{2})}{\int_{0}^{\infty}
	dx~f(x)\exp(-\frac{Q_{b}}{2})}\label{eqn:peak}
\end{eqnarray}   
Since the detailed derivation of the above probability and
explanations of the notation are complicated, we omit them here (see
YNG).  By spatially averaging eq.(\ref{eqn:peak}) in the region of
$M_{2}$, we obtain $P(M_{1},M_{2})$ as follows:
\begin{equation}   
P(M_{1},M_{2})=\frac{\int_{0}^{R_{2}}P(r,M_{1},M_{2}|\mbox{peak})   
	4\pi r^{2}dr}{\int_{0}^{R_{2}}4\pi r^{2}dr}\label{eqn:mean},   
\end{equation}   
where $R_{2}$ means the radius of the region $M_{2}$,
$R_{2}=(3M_{2}/4\pi\bar\rho)^{(1/3)}$.  The cumulative multiplicity
functions, which will be defined in the next section, are estimated by
using eq.(\ref{eqn:peak}) and are shown by the short-dashed lines in
Fig.3 in the cases of the power spectrum $P(k)\propto k^{n}$ with
$n=0$ and $-2$, respectively.  We call this formula the YNG formula
hereafter.
   
\section{RESULTS}  

\subsection{Cumulative multiplicity functions}  
  
We calculate the cumulative multiplicity function $P(\geq M)$, which
is defined as the mass fraction of objects whose mass is larger than a
mass $M$ to the total mass (in a fair sample of the Universe),
by following the various
methods which are mentioned in \S \S 2 and 3.  The multiplicity
function $P(M)d\ln M$ (which we define in a logarithmic interval of mass)
and the cumulative multiplicity function are related to the mass
function $n(M)dM$ as follows:
\begin{eqnarray}  
P(M)d\ln M&\equiv&\frac{d\rho}{\bar\rho}= 
\frac{M^{2}n(M)}{\bar\rho}d\ln M,\\  
P(\geq M)&=&\int_{M}^{\infty}P(M)\frac{dM}{M}.  
\end{eqnarray}  
We assume that the linear density fluctuations obey a random Gaussian
distribution with a power-law power spectrum $P(k)\propto k^{n}$,
where $n=0$ and $-2$.  We consider only the Einstein-de Sitter universe
($\Omega=1, \Lambda=0$) in this paper.  In the MCM, we take the box
size $L=128$.  $M_{*}$ is defined as $\sigma(M_*)=1$ (see eq.(3)) and
here the mass in the block with the mass scale $M_*$ is assigned to
eight cells.  In the Block model, the largest box size is assigned to
the mass scales $M_{0}=10^{5}M_{*}$ in the case of $n=0$ and
$10^{7}M_{*}$ in the case of $n=-2$.  We consider block sizes
with $M_i=M_0/2^i(i=1, 2, \cdots, N=27)$.
  
In Figs.3(a) and (b), we show the cumulative multiplicity functions
derived by the MCM, the YNG formula, the PS formula and the Block
model for the case of $n=0$ and $n=2$, respectively.

\begin{figure}
\epsfxsize=8cm
\epsfbox{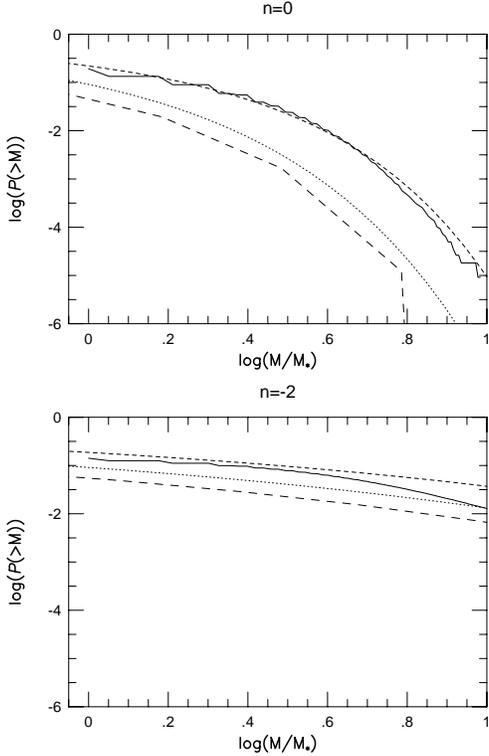}
\caption{Cumulative multiplicity function: (a)spectral index n=0,
(b)n=-2.  The solid lines, the short-dashed lines, the dotted lines
and the long-dashed lines show the cumulative multiplicity functions
given by the MCM averaged over four realisations, the YNG formula, the
PS formula and the Block model averaged over five realisations,
respectively.}
\end{figure}

The multiplicity functions of the MCM are shown only in the mass range
($0 \leq \log(M/M_{*}) \la 1$) which corresponds to the range from
eight cells to about 100 cells.  The reason is as follows: On the
smallest mass scale (one cell scale) in the numerical calculations,
the power of the density fluctuations on the one cell scale cannot be
correctly produced in agreement with the theoretically estimated power
of the scale-free mass spectrum $\sigma(M)\propto M^{-(3+n)/6}$ due to
the finiteness of the cell size in the numerical calculations.  On the
larger mass scales, larger haloes than about 100 cells have various
shapes, rather than the spherical shape, so the assumption of
spherically symmetric collapse fails.
Hence, the cumulative multiplicity function given by different method
on these scales cannot be directly computed without clarifying the
identification of the isolated haloes.  Moreover, since there are few
haloes on the scales larger than about 100 cells in the numerical
calculations, the error in the number of haloes increases.  Therefore,
we show here the cumulative multiplicity function only on the mass
range from eight cells to about 100 cells ($0\leq\log(M/M_*)\la 1$).

It is found that the multiplicity functions given by the MCM are well
fit by those given by the YNG formula.  In the case of $n=0$, the
agreement is good while in the case of $n=-2$, there is a little
deviation on the mass scale $\log(M/M_{*})\ga 0.8$ because of the
numerical errors mentioned above.  It is also found that the PS
formula and the Block model are also in agreement with each other in
the case of $n=-2$ rather than in the case of $n=0$.  Note that on the
galaxy scales the spectral index $n$ is nearly equal to $-2$ in the
CDM model.  Then in the CDM model the both functions are consistent
with each other (Cole \& Kaiser 1988).  It must be noticed that the
agreement of the Block model on the PS formula depends on the power
spectrum.  On the other hand, the function of the MCM deviates from
those of the Block model and the PS formula in which the spatial
correlations of the density fluctuations are not explicitly taken into
account.  From these results, we can conclude that the MCM naturally
and correctly takes into consideration the spatial correlations and
the deviation of the multiplicity function of the MCM from those of
the Block model and the PS formula results from the effect of the
spatial correlations of the density fluctuations.  Furthermore, we
will show in the next subsection that the multiplicity function given
by the Block model can be reproduced by using the Jedamzik formula
without consideration of the spatial correlations and thereby
demonstrate that the difference between the multiplicity functions
given by the Block model and the MCM results from the effect of the
spatial correlations.

\subsection{Block model}  
Here we analytically reproduce the multiplicity function given by the 
Block model by using the Jedamzik formula. 
  
Since we consider the discrete mass of the blocks ($M_{i}=M_{0}/2^i$)
in the Block model, the density contrast in the blocks which are
identified as isolated collapsed haloes is generally greater than
$\delta_{c}$.  We cannot recognise the just collapsed halo whose
density contrast is precisely $\delta_c$ if we follow the procedure of the
Block model.  Therefore, the conditional probability $P(M_1, M_2)$ in
this case must be approximately expressed as:
\begin{eqnarray}  
	P(M_{1},M_{2})&=&p(\delta_{M_{1}}\geq\delta_{c}|\delta_{M_{2}}  
	\geq\delta_{c})\nonumber\\  
	&=&\frac{p(\delta_{M_{1}}\geq\delta_{c},\delta_{M_{2}}  
	\geq\delta_{c})}{p(\delta_{M_{2}}\geq\delta_{c})}\nonumber\\  
\lefteqn{=\frac{\int_{\nu_{1c}}^{\infty}d\nu_{1}  
	\int_{\nu_{2c}}^{\infty}d\nu_{2}\exp\left\{-\frac  
	{(\nu_{1}-\epsilon'\nu_{2})^{2}}  
	{2(1-\epsilon'^{2})}-\frac{\nu_{2}^{2}}{2}\right\}}  
	{\sqrt{2\pi(1-\epsilon'^{2})}\int_{\nu_{2c}}^{\infty}d\nu_{2}  
	\exp\left\{-\frac{\nu_{2}^{2}}{2}\right\}}.}\label{eqn:bl}  
\end{eqnarray}  
It must be noticed here that we must consider the case of
$\delta_{M_2} \geq \delta_c$ instead of the case of
$\delta_{M_2}=\delta_c$ which appeared in eq.(11).  In this case, of course,
the spatial correlation is not taken into consideration.  The detailed
derivation of the above probability and notation are explained in
Appendix A.  Inserting the above conditional probability, eq.(17),
into eq.(10), we can estimate the multiplicity function in this case.
  
\begin{figure}
\epsfxsize=8cm
\epsfbox{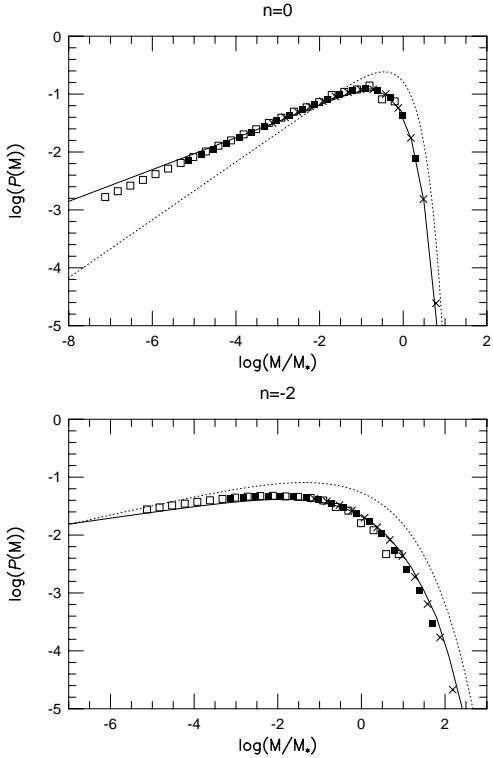}
\caption{Block Model: (a)n=0, (b)n=-2.  Crosses, solid squares and open
squares show the multiplicity functions averaged over five realisations
of the Block model with the largest box size $M_{0}=10^5M_{*},
10^3M_{*}$ and $10M_{*}$ in the case of $n=0$ and $10^{7}M_{*},
10^{5}M_{*}$ and $10^{3}M_{*}$ in the case of $n=-2$,
respectively.  The solid lines and the dotted lines show the predictions
by using the Jedamzik formula and the PS formula, respectively.}
\end{figure}

In Fig.4, we show the multiplicity functions given by the above
procedure (solid line) and the Block model (cross, solid square and open
square).  The cross, solid square and open marks in the Block model
mean that the masses of the largest block $M_{0}$ equal $10^5M_{*},
10^3M_{*}$ and $10M_{*}$, respectively.  Here in the Block model, the
multiplicity function over a large dynamic range in mass can be produced by
combining the functions derived by the Block model with
$M_0=10^5M_{*}, 10^3M_{*}$ and $10M_{*}$ in the case of $n=0$ and
$10^{7}M_{*}, 10^{5}M_{*}$ and $10^{3}M_{*}$ in the case of $n=-2$ and
block sizes with $M_i=M_0/2^i(i=1,2, \cdots 27)$ in each case.  We
find that the multiplicity functions given by the Block model and the
analytical formula derived from the Jedamzik formula fit each other well.  We
believe that the difference between the cumulative functions given
by the Block model and the MCM mainly results from the effect of
the spatial correlations while the difference between the functions
given by the Block model and the PS formula results from the
difference in the identification of isolated collapsed objects.

\subsection{Overlapping effect}  
  
In taking into consideration the finite size of the haloes, it appears
necessary to consider the serious problem of the spatial overlapping
of the dark haloes.  For the overlapping criterion we must consider
how we can identify the number and sizes of the haloes when some
haloes overlap.  In this subsection, we show the effect of the
overlapping criterion (\S 2.2,(b)) on the multiplicity function.  The
overlapping criterion adopted in RT is that the investigating region
can collapse when the overlapping region has a larger mass than half
of the lesser of the masses of the halo and the investigating region.
Here we quantify the overlapping criterion by defining a parameter,
$x$, as the ratio of the mass of the overlapping region to the lesser
of the masses of the halo and the investigating region.  Then, RT's
criterion corresponds to $x=1/2$.  By changing the value of $x$ we
show the effect of overlapping on the multiplicity function.
  
In Fig.5, we show the multiplicity functions on scales above eight
cells in the cases that $x=1/2$, $1/4$ and $1/8$.  As the value of $x$
decreases, the multiplicity function increases on the larger mass
scales because it becomes easy for larger blocks to collapse.

\begin{figure}
\epsfxsize=8cm
\epsfbox{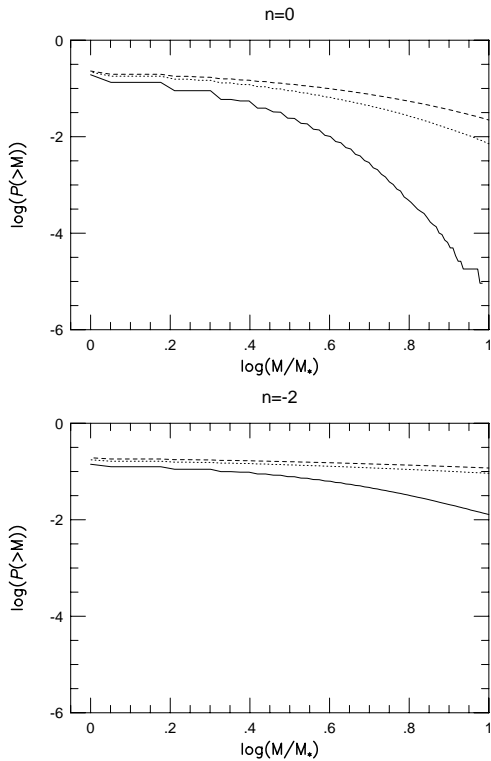}
\caption{Effects of overlapping: (a)n=0, (b)n=-2.  The solid
lines, dotted lines and dashed lines show the multiplicity functions
averaged over four realisations given by the MCM with the
overlapping criterion defined by $x=1/2, 1/4$ and $1/8$, respectively.}
\end{figure}
  
We find that the multiplicity function on the larger mass scales
changes according to the value of $x$.  Note that this change is
shown quantitatively in Fig. 5 only on scales smaller than about 100
cells because of the uncertainties due to the numerical errors on
scales larger than 100 cells which are mentioned in \S 4.1.  We
consider the trend of the multiplicity function on large mass scales
to increase with decreasing $x$ on scales larger than 100 cells to
also be significant.

So, the overlapping effect is a serious problem for constructing
merger trees of dark haloes, particularly on mass scales larger than
about 100 cells.

\section{CONCLUSIONS \& DISCUSSIONS}  
  
In this paper, we have shown that the multiplicity functions given by the 
MCM are consistent with those given by the YNG formula.  However, the 
functions of the MCM do not fit those given by the PS formula and the 
Block model.  This fact means that the MCM includes the information of 
the spatial correlations in the density fluctuation field, but the PS 
formula and the Block model do not include it.  Thus we believe that 
the effects of the spatial correlations affect the merger trees as 
well as the mass functions and it is important to take into account 
the spatial correlations in the merger tree models. 
  
Here it must be noted that the reason why the multiplicity functions 
given by the MCM are well fit by those given by the YNG formula in the case 
that the overlapping criterion has the value of $x=1/2$ is as follows. 
When obtaining the multiplicity function~from eqs.(\ref{eqn:peak}) and 
(\ref{eqn:mean}), we integrate the $P(r,M_{1},M_{2})$ with respect to 
$r$~from $0$ to $R_{2}$ in eq.(\ref{eqn:mean}).  This corresponds to 
the case that $x=1/2$.  Of 
course, the adoption of the integration interval 
$[0,R_{2}]$ is, strictly speaking, only justified 
for spherically symmetric collapse.
In fact, 
however, the effects of non-spherical collapse or tidal 
interaction among haloes may be important.  We hypothesise that
changing the value of $x$ might be enough to account for these effects.
    
We also analytically produce multiplicity functions which fit 
those given by the Block model, by using the Jedamzik formula.  When 
we reproduce the multiplicity functions given by the Block model, the 
conditional probability eq.(\ref{eqn:bl}) is adopted instead of 
eq.(\ref{eqn:jed}) in the PS formula.  The difference from the PS 
formalism is that the density contrast of the isolated object is not 
only equal to $\delta_{c}$ but also greater than $\delta_{c}$.  This 
difference results from the discreteness of block mass in the 
Block model.  It must be noticed that the conditional probability 
eq.(\ref{eqn:bl}) does not include the effect of the spatial 
correlations of the density fluctuations.  We find that the 
multiplicity functions reproduced in this way are consistent with those 
given by the Block model.  This result shows that the Block model 
does not include the spatial correlations of the density fluctuations 
and so the multiplicity function given by the Block model is different 
 from those given by the MCM. 
  
Moreover, we have shown that when the overlapping criterion is
changed, the multiplicity function is affected, especially on larger
mass scales.  As the value of the overlapping parameter, $x$ (defined
as the ratio of the mass of the overlapping region to the lesser of
the masses of the halo and the investigating region) decreases, the
number density of high mass haloes increases because it becomes easy
for larger blocks to collapse.

We believe that the MCM is a powerful tool for constructing merger
trees of dark haloes because it includes the information of spatial
correlations naturally and correctly as shown in this paper.  In the
MCM, the number of high mass haloes increases, relative to the results
of previous works, {\it e.g.}, Kauffman et al. (1993) and Cole et
al. (1994), so the number of giant, red elliptical galaxies may
increase.  In the MCM, however, a serious problem appears, {\it i.e.},
it is found that in taking into consideration the overlapping of dark
haloes of non-zero size, the multiplicity function is sensitive to the
choice of the overlapping criterion.  Therefore, we must consider a
more realistic overlapping criterion for constructing merger trees.
We do not know {\it a priori} the value of the overlapping
parameter. Furthermore the value of the overlapping parameter may not
be constant as time passes.  It might be a function of halo mass, of
separation between a halo and a block, or some other parameter.  It is
a very difficult problem to identify the mass and size of the
collapsed halo when several haloes are overlapped.  However, it is
very important to precisely identify the collapsed haloes for
constructing the merger trees of haloes with good accuracy.  As
mentioned in $\S 1$, the merger tree of the dark haloes greatly
affects the various important processes for the formation and
evolution of galaxies, such as the thermal history of gases, merging
between galaxies and so on.  Hence, it is necessary and important to
understand how the overlapping criterion influences these processes.
Furthermore, we assume spherically collapse of the dark haloes.  But,
in fact, there is the possibility that the collapsed objects have
filament or sheet structures.  We must also deal with these cases in
order to correctly estimate the merger trees.  We are now investigating
the merger tree models in which the effect of non-spherical collapse
is included and the identification of the collapsed haloes is well
defined.

\section*{ACKNOWLEDGMENTS}    
We wish to thank Misao Sasaki and Satoru Ikeuchi for useful
suggestions, and Taihei Yano for stimulated discussions.  We are
grateful to Boudewijn F. Roukema for reading carefully the manuscript
of our paper.  This work was supported in part by Research Fellowships
of the Japan Society for the Promotion of Science for Young Scientists
and in part by the Grant-in-Aid No.06640352 for the Scientific
Research Fund from the Ministry of Education, Science and Culture of
Japan.  The calculations were performed in part on a VPP300 at the
Astronomical Data Analysis Center in National Astronomical
Observatory.

\appendix   
\section{MASS FUNCTION OF BLOCK MODEL}   
   
As mentioned in \S 3, we show the analytic formula which reproduces  
the mass functions given by the Block model.  
   
The analytic formula is approximately expressed as follows.  In YNG
and PS, it is assumed that the density contrast $\delta_{M_{2}}$ of
the isolated object which has collapsed with mass $M_2$ is just
$\delta_{c}=1.69$ (see eq.(13)).  In the Block model, however, this
condition should be changed to $\delta_{M_{2}}\geq\delta_{c}$ because
in general we can not see the block which has just collapsed and its
density contrast is just $\delta_c$ due to the discreteness of the
mass scales of the blocks with $M_{i+1}=M_{i}/2$ in the Block model.
   
So the probability should be changed as follows:   
\begin{eqnarray}   
P(M_{1},M_{2})&=&p(\delta_{M_{1}}\geq\delta_{c}|\delta_{M_{2}}\geq\delta_{c})
	\nonumber\\
	&=&\frac{p(\delta_{M_{1}}\geq\delta_{c},\delta_{M_{2}}\geq\delta_{c})}
	{p(\delta_{M_{2}}\geq\delta_{c})}\label{eqn:bayes}.
\end{eqnarray}   
In order to estimate the probability on the right hand side of  
eq.(\ref{eqn:bayes}), we have to consider the two-variables Gaussian  
distribution function.  The probability of $m$-variables Gaussian is  
generally (see BBKS)  
\begin{equation}   
p({\bf V})d{\bf V}=\frac{\exp(-Q/2)}{\sqrt{(2\pi)^{m}\det({\bf   
M})}}d{\bf V},   
\end{equation}   
where   
\begin{eqnarray}   
Q&=&{\bf VM}^{-1}{\bf V}^{T},\\   
{\bf V}&=&(y_{1},y_{2},\ldots,y_{m}),\\   
M_{ij}&=&\langle(y_{i}-\langle y_{i}\rangle)(y_{j}-\langle   
y_{j}\rangle)\rangle   
\end{eqnarray}   
and $y_{i} (i=1,2,\ldots,m)$ is the Gaussian random variables.  We use  
the angle brackets $\langle \rangle$ as the ensemble average of the  
universe, but in practice, assuming homogeneity and ergodicity in  
space, we can take it as the spatial average. So $\langle  
y_{i}\rangle$ corresponds to the spatial average of $y_{i}$, and  
$M_{ij}$ is covariance between $y_{i}$ and $y_{j}$.  In this case, we  
may consider the two variables, $\delta_{M_{1}}$ and $\delta_{M_{2}}$,  
so the covariance matrix ${\bf M}$ is written as follows:  
\begin{eqnarray}   
{\bf M}=\left(   
	\begin{array}{cc}   
	\langle\delta_{M_{1}}^{2}\rangle & \langle\delta_{M_{1}}   
						\delta_{M_{2}}\rangle \\   
 	\langle\delta_{M_{1}}\delta_{M_{2}}\rangle &    
				\langle\delta_{M_{1}}^{2}\rangle    
	\end{array}   
\right).   
\end{eqnarray}	   
Note that $\langle\delta_{M_{1}}^{2}\rangle$ and  
$\langle\delta_{M_{1}}^{2}\rangle$ are variances, $\sigma^{2}(M_{1})$  
and $\sigma^{2}(M_{2})$, respectively, and  
$\langle\delta_{M_{1}}\delta_{M_{2}}\rangle$ is a cross correlation at  
the same point.  
   
Here, we normalize the density contrast and the cross correlation,   
\begin{equation}   
\nu_{i}\equiv\frac{\delta_{M_{i}}}{\sigma(M_{i})},\quad  
\epsilon\equiv\frac{\sigma^{2}_{h}}{\sigma(M_{1})\sigma(M_{2})},\quad  
\sigma^{2}_{h}\equiv\langle\delta_{M_{1}}\delta_{M_{2}}\rangle.  
\end{equation}   
Using the above notation, we obtain the two variables Gaussian   
distribution function,   
\begin{eqnarray}
\lefteqn{p(\nu_{1},\nu_{2})d\nu_{1}d\nu_{2}=}\nonumber\\
&&\frac{1}{2\pi\sqrt{1-\epsilon^{2}}}
\exp\left(-\frac{
(\nu_{1}-\epsilon\nu_{2})^{2}}{2(1-\epsilon^{2})}-\frac{\nu_{2}^{2}}{2}\right)
d\nu_{1}d\nu_{2}.
\end{eqnarray}   
So we obtain  
\begin{eqnarray}   
\lefteqn{p(\delta_{M_{1}}\geq\delta_{c},\delta_{M_{2}}\geq\delta_{c})=
p(\nu_{1}\geq\nu_{1c},\nu_{2}\geq\nu_{2c})}\nonumber\\
&&=\frac{1}{2\pi\sqrt{1-\epsilon^{2}}}\nonumber\\
&&\quad\times\int_{\nu_{1c}}^{\infty}\int_{\nu_{2c}}
^{\infty}\exp\left(-\frac{
(\nu_{1}-\epsilon\nu_{2})^{2}}{2(1-\epsilon^{2})}-\frac{\nu_{2}^{2}}{2}\right)
d\nu_{1}d\nu_{2}\label{eqn:two},
\end{eqnarray}   
where $\nu_{ic}$ is $\delta_{c}/\sigma(M_{i})$.  Furthermore, dividing   
the above equation by the integration of one variable Gaussian   
distribution,   
\begin{eqnarray}   
\lefteqn{p(\delta_{M_{2}}\geq\delta_{c})=p(\nu_{2}\geq\nu_{2c})}\nonumber\\
&&=\frac{1}{\sqrt{2\pi}}\int_{\nu_{2c}}^{\infty}\exp\left(-\frac{\nu_{2}^{2}}
{2}\right)d\nu_{2},
\end{eqnarray}   
we obtain the $P(M_{1},M_{2})$.   
   
Next, we estimate the normalized cross correlation function   
$\epsilon$.  In eqs.(\ref{eqn:vari}) and (\ref{eqn:sharp}), the   
variance $\sigma_{i}$ with the mass $M_{i}$ is given by   
\begin{eqnarray}   
\sigma_{i}^{2}&=&\frac{4\pi   
	V}{(2\pi)^3}\int_{0}^{\infty}W_{i}^{2}(k)P(k)k^{2}dk,   
\end{eqnarray}   
and the variance of the additional density contrast $\Sigma$ is given by   
\begin{eqnarray}   
\Sigma_{i}^{2}&=&\sigma_{i}^{2}-\sigma_{i-1}^{2},~~~i\geq 1.   
\end{eqnarray}   
Because we give the additional density contrast independently of  
another density contrast with the other mass scale, the variance of  
the additional density contrast must depend on the power spectrum   
only at the interval of the wavenumber $[k_{i+1},k_{i}]$, corresponding  
to the mass scales $[M_{i},M_{i+1}]$.  So we should adopt the  
sharp-$k$ space filter as the window function $W_{i}(k)$ defined by,  
\begin{equation}   
  W_{i}(k)=\left\{   
  \begin{array}{ll}   
    1,&\quad k\leq k_{c}(M_{i})\\   
    0,&\quad k>k_{c}(M_{i}),   
  \end{array}\right.   
\end{equation}   
   
In this case, the variance and the cross correlation are   
\begin{eqnarray}   
\sigma_{i}^{2}&=&\frac{4\pi   
	V}{(2\pi)^3}\int_{0}^{k_{i}}P(k)k^{2}dk,\\   
\sigma_{h}^{2}&=&\frac{4\pi   
	V}{(2\pi)^3}\int_{0}^{\infty}W_{1}(k)W_{2}(k)P(k)k^{2}dk\nonumber\\   
&=&\frac{4\pi   
	V}{(2\pi)^3}\int_{0}^{k_{2}}P(k)k^{2}dk\nonumber\\   
&=&\sigma_{2}^{2}   
\end{eqnarray}   
where $i=1,2$ and $M_{1}\leq M_{2}$, {\it i.e.}, $k_{1}\geq k_{2}$.   
So the normalized cross correlation function $\epsilon$ is   
\begin{equation}   
\epsilon=\frac{\sigma_{h}^{2}}{\sigma_{1}\sigma_{2}}=\frac{\sigma_{2}}   
{\sigma_{1}}.\label{eqn:eps}   
\end{equation}   
Substituting the above eq.(\ref{eqn:eps}) into eq.(\ref{eqn:two}), we
obtain
\begin{eqnarray}   
\lefteqn{p(\delta_{M_{1}}\geq\delta_{c},\delta_{M_{2}}\geq\delta_{c})
=\frac{1}{2\pi\sigma_{2}\sigma_{\mbox{sub}}}}\nonumber\\
&&\times\int_{\delta_{c}}^{\infty}\int_{\delta_{c}}
^{\infty}\exp\left(-\frac{
(\delta_{1}-\delta_{2})^{2}}{2\sigma_{\mbox{sub}}^{2}}
-\frac{\delta_{2}^{2}}{2\sigma_{2}}\right) d\delta_{1}d\delta_{2},
\end{eqnarray}   
where   
\begin{equation}   
\sigma_{\mbox{sub}}^{2}=\sigma_{1}^{2}-\sigma_{2}^{2}.   
\end{equation}   
Therefore the conditional probability $P(M_{1},M_{2})$ is    
\begin{eqnarray}   
\lefteqn{P(M_{1},M_{2})=N^{-1}\frac{1}{2\pi\sigma_{2}\sigma_{\mbox{sub}}}}
\nonumber\\
&&\times\int_{\delta_{c}}^{\infty}\int_{\delta_{c}}
^{\infty}\exp\left(-\frac{
(\delta_{1}-\delta_{2})^{2}}{2\sigma_{\mbox{sub}}^{2}}
-\frac{\delta_{2}^{2}}{2\sigma_{2}}\right) d\delta_{1}d\delta_{2},
\end{eqnarray}
where
\begin{equation}
N=\frac{1}{\sqrt{2\pi}\sigma_{2}}\int_{\delta_{c}}^{\infty}\exp\left(
-\frac{\delta_{2}^{2}}{2\sigma_{2}}\right)d\delta_{2}.
\end{equation}
Note that this probability is the same as eq.(8) in Jedamzik (1995). 

\bsp

\begin{thebibliography}{}   
\bibitem{}Bardeen J.M., Bond J.R., Kaiser N., Szalay A.S., 1986, ApJ,
304, 15(BBKS)
\bibitem{}Bond J.R., Cole S., Efstathiou G., Kaiser N., 1991, ApJ,
379, 440
\bibitem{}Bower R.J., 1991, MNRAS, 248, 332
\bibitem{}Cole S., 1989, ApJ, 367, 45
\bibitem{}Cole S., Aragon-Salamanca A., Frenk C.S., Navarro J.F., Zepf
S.E., 1994, MNRAS, 271, 781
\bibitem{}Cole S., Kaiser N., 1988, MNRAS, 233, 637
\bibitem{}Epstein R.I., 1983, MNRAS, 205,207   
\bibitem{}Evrard A.E., Summers F.J., Davis M., 1994, ApJ, 422, 11   
\bibitem{}Jedamzik K., 1995, ApJ, 448, 1   
\bibitem{}Katz N., 1992, ApJ, 391, 502   
\bibitem{}Katz N., Gunn J.E., 1991, ApJ, 377, 365   
\bibitem{}Kauffman G., White S.D.M., 1993, MNRAS, 261, 921 (KW)  
\bibitem{}Kauffman G., White S.D.M., Guiderdoni B., 1993, MNRAS, 271, 781   
\bibitem{}Lacey C.G., Cole S., 1993, MNRAS, 262, 627   
\bibitem{}Lacey C.G., Silk J., 1991, ApJ, 381, 14   
\bibitem{}Lacey C.G., Silk J., 1993, ApJ, 402, 15   
\bibitem{}Navarro J.F., White S.D.M., 1993, ApJ, 265, 271   
\bibitem{}Peacock J.A., Heavens A.F., 1985, MNRAS, 217, 805   
\bibitem{}Peacock J.A., Heavens A.F., 1990, MNRAS, 243, 133   
\bibitem{}Peebles P.J.E., 1993, The Principles of Physical   
Cosmology, Princeton: Princeton Univ. Press   
\bibitem{}Press W.H., Schechter P., 1974, ApJ, 387, 47(PS)   
\bibitem{}Rees M.J., Ostriker J.P., 1977, MNRAS, 267, 1020   
\bibitem{}Rodrigues D.D.C., Thomas P.A., 1995, submitted to MNRAS,   
astro-ph/9511018   
\bibitem{}Steinmetz M., Muller E., 1994, A\&A, 281, L97
\bibitem{}White S.D.M., Frenk C.S., 1991, ApJ, 379, 25   
\bibitem{}White S.D.M., Rees M.J., 1978, MNRAS, 183, 341   
\bibitem{}Yano T., Nagashima M., Gouda N., 1996, ApJ, 466, 1(YNG)   
\end{thebibliography}
\end{document}